\documentclass[12pt,a4paper]{article}
\usepackage{jheppub}
\usepackage{chemarrow}
\usepackage{amsmath,amssymb,amsfonts,pstricks,mathtools}
\usepackage{float}
\usepackage{subfig}
\usepackage{scalefnt}
\usepackage{wrapfig}
\usepackage{fancybox}
\usepackage{ifsym}
\newcommand{\bea}{\begin{eqnarray}\displaystyle}
\newcommand{\eea}{\end{eqnarray}}

\newlength{\arrow}
{\setlength{\fboxsep}{15pt}
\setlength{\mylength}{\linewidth}%
\addtolength{\mylength}{-2\fboxsep}%
\addtolength{\mylength}{-2\fboxrule}%
\Sbox
\minipage{\mylength}%
\setlength{\abovedisplayskip}{0pt}%
\setlength{\belowdisplayskip}{0pt}%
\equation}%
{\endequation\endminipage\endSbox
\[\fbox{\TheSbox}\]}

\def\Dsl{\,\raise.15ex\hbox{/}\mkern-13.5mu D}

\begin{document}
\title{Non-Unitary Holography}

\author{Cumrun Vafa}

\affiliation{Jefferson Physical Laboratory, Harvard University, Cambridge, MA 02138, USA}

\abstract{We propose gauge theory/gravity duality involving conformal theories
 based on $U(N+k|k)$ gauge groups.  We show that to all orders in $1/N$ these non-unitary theories based
on supergroups are indistinguishable from the corresponding unitary theories where the gauge group is
replaced by $U(N)$.  This leads to non-unitary gravity duals which to all orders in $1/N$ are indistinguishable
from their unitary cousins.  They are distingushed by operators whose correlation functions differ by $O(e^{-aN})$.
The celebrated type IIB on $AdS^5\times S^5$ and M-theory on $AdS^4\times S^7$ fall in this class and thus seem
to also admit non-unitary 
non-perturbative completions.
It is tempting to conjecture that this setup may provide a non-unitary model for black hole evaporation. }

\maketitle

\section{Introduction}
Holographic duality between conformal theories and gravity \cite{malda,gkp,wittenads} has led
to deep insights about both gauge theory and gravity.  The interest for the gauge theory
side involves being able to solve the large $N$ strong coupling regime of the theory using
semi-classical gravity.  The interest from the gravity side involves being able to define quantum gravity
on certain spaces using the better understood gauge theory.  In particular the duality between quantum
gravity and unitary gauge theories  leads to the prediction that quantum gravity, including
black hole evaporation process, is unitary.  This is indeed a remarkable achievement.

On the other hand, we also know that  non-trivial non-unitary conformal theories exist and
have equally rich structure.
The simplest class of such models involves the Yang-Lee edge singularity which corresponds
to 2d Ising model with imaginary magnetic field,  and has been identified as a $(2,5)$ member \cite{cardy}
of the $(p,q)$ minimal  non-unitary\footnote{There are also non-unitary minimal models for superconformal theories
in 2d.}
conformal filed theories \cite{bpz}. 
It corresponds to the fixed point of a real scalar field in 2d with $ig \phi^3$ potential \cite{fisher}.  This is a non-unitary theory
as the action is not real.\footnote{As is usually the case in non-unitary theories we can make the action real
at the cost of making the kinetic and potential energy unbounded from below by redefining $\phi \rightarrow i\phi$.
 This example illustrates the point that theories whose path-integrals are difficult to make sense out of
due to convergence issues,
can nevertheless lead to consistent, rich albeit non-unitary quantum field theories.}
Another class of non-unitary 2d CFT's involves WZW models with supergroups as their target space
(see \cite{schomerus1} and references therein) and
their 3d Chern-Simons counterparts \cite{wittennew}.
Indeed these models are interesting also from the perspective of describing the worldsheet theory associated
to various $AdS$ backgrounds that arise in holographic duality (see e.g. \cite{bvw}).
Other applications of non-unitary conformal theories involve their application in conformal turbulence
\cite{polyakov}.  Furthermore, non-unitary
conformal theories corresponding to sigma models on supermanfiolds play a role in the context of mirror symmetry in the `non-geometric' cases \cite{sethi}.  They also appear in the worldsheet description of twistor string \cite{wittwist}.
Moreover at least in some examples they can also be extended from Euclidean to Lorentzian signature \cite{mo}.

Given the rich structure enjoyed by non-unitary conformal theories it is natural to ask whether they also
lead to interesting holographic duals.  It is the purpose of this paper to give an affirmative answer to this
question.  The class of theories we consider is nicely exemplified by ${\cal N}=4$ SYM theory based on $U(N+k|k)$ supergroup.
It is natural to expect this theory to exist and to be a non-unitary conformal theory based on the high degree of supersymmetry.
However in exploring their gravity duals one is led to the unexpected (and maybe even unwanted) situation where the well known holographic
duals to ${\cal N}=4$ $U(N)$ Yang-Mills theory, such as $AdS^5\times S^5$, are indistinguishable
from their non-unitary cousins to all orders in $1/N$, i.e., to all orders in string perturbation theory!
What distinguishes them is that the relation between Casimirs in the $U(N)$ theory, the `stringy
exclusion principle' \cite{ms}
are not satisfied by the $U(N+k|k)$ theory.  For example if one considers operators of the form
$${\cal O}= tr F^{N+1}-\sum_{\sum a_i=N+1}C_{a_i}\prod_i trF^{a_i}$$
where one  chooses the coefficients  $C_{a_i}$  so that operator is zero in the $U(N)$ theory due to Casimir relations,
the same operator (with traces interpreted as supertraces) is non-trivial in the $U(N+k|k)$ theory.  However, correlations involving it
will be non-perturbative in $1/N$:
 $$\langle {\cal O} A B C...\rangle \sim  O(e^{-aN}).$$
In other words, their difference is not detectable to all orders in the $1/N$ expansion.

The organization of this paper is as follows:   In section 2 we review some facts about super Lie algebras.
In section 3 we propose our non-unitary holographic duals.
In section 4 we discuss supermatrix models as a toy model as well as a way to compute
certain supersymmetric amplitudes in these theories which leads to insights about the structure of
these theories.   In particular we prove that certain correlations functions differ between ${\cal N}=4$  supersymmetric $U(N+1|1)$ theory 
and $U(N)$ theory by non-perturbative terms in $N$.
In section 5 we conclude
by raising some open questions.

\section{$U(N+k|k)$ supergroups and their Casimirs}

$U(N+k|k)$ is defined as the group that preserves the natural norm in the ${\bf C}^{N+k|k}$ superspace 
given for each element $v=(z,\theta) \in {\bf C}^{N+k|k}$ by
$$|v|^2=\sum_{i=1}^{N+k} |z_i|^2+\sum_{j=1}^k {\overline \theta_j} \theta_j.$$
Let $M$ be an element in the Lie algebra of this supergroup viewed as an $(N+2k) \times( N+2k) $ matrix. The upper
$(N+k)\times (N+k)$ block and the lower $k\times k$ block is bosonic and the other elements are fermionic.  
 One defines a supertrace as follows:  Let $x_i$ be the first $N+k$ diagonal element and $y_j$ the
last $k$ diagonal elements.  The supertrace is defined by
$$Str M=\sum_{i=1}^{N+k} x_i -\sum_{j=1}^k y_j$$
Note that we have to take the supertrace to get a group invariant.
In particular the usual trace is {\it not} group invariant.  In the physical
context, this means that, unlike supersymmetric traces, where we have the option
of inserting or not inserting $(-1)^F$, for theories involving gauge invariance for supergroups,
we have no such option: we need to take supertrace.

The invariant Casimirs associated to $M$ are given by supertraces of arbitrary powers
of $M$:
$$O_p=Str M^p$$
There are $N+2k$ independent Casimirs.  However, unlike the $U(N)$ case
the supertrace of higher powers $p>N+2k$ of $M$ are rational rather than polynomial
combinations of lower Casimirs. Viewing $M$ as a diagonal matrix, the ring generated by the Casimirs of $M$ can be identified
\cite{sergeev1,sergeev2} with functions $f(x_i,y_j)$ which is invariant under the separate
permutation of $x$'s and $y$'s among
themselves and satisfies 
$${\partial f\over \partial x_i}+{\partial f\over \partial y_j}=0\bigg|_{x_i=y_j}$$

As an application of this last statement that we shall use later in this paper note that
$$f=\prod_{i,j} (x_i-y_j) $$
satisfies these properties.  Let us consider the special case of $U(N+1|1)$.
Then since $f$ is an invariant of degree $N+1$ we have
$$f=\prod_{i,j}(x_i-y)=\sum_{\sum a_i=(N+1)} C_{a_i} \prod_i Str M^{a_i}$$
Let us consider this identity restricted to the $U(N)\subset U(N+1|1)$ subalgebra, i.e.,
in the special case where $x_{N+1}=y=0$.  In this case
the supertrace reduces to ordinary trace of an $N\times N$ matrix.  On the other hand
the left hand side vanishes in this limit.  Since $x_1,...,x_N$ are generic numbers,
this can only be possible if the $C_{a_i}$ are the same as the coefficients appearing
in the Casimir relation of $U(N)$ expressing the trace of $N+1$-st power in terms
of the lower powers.   In other words viewing $U(N)\subset U(N+1|1)$ the
above invariant vanishes on the $U(N)$ subalgebra due to Casimir relations, but not on the $U(N+1|1)$. 
 We will
use this fact when we do explicit computations in the context of the $U(N+1|1)$ matrix models
in section 4. 

Consider for example $N=1$ case, corresponding to $U(2|1)$.  In that case we have
$$f=(x_1-y)(x_2-y)={-1\over 2}\big[Str M^2 -(Str M)^2\big]$$
Note that the right hand side would be zero for the $U(1)$ case.  Similarly for the $N=2$ case we have
$$f=(x_1-y)(x_2-y)(x_3-y)={1\over 3}\big[Str M^3-{3\over 2}Str M Str M^2 +{1\over 2} (Str M)^3\big]$$
Again it is easy to check that the right hand side is zero for the $U(2)$  case, but not for the $U(3|1)$.

\section{Non-unitary holography}
In this section we discuss our conjecture about non-unitary holography.  

We first show a correspondence between two theories:  Consider any gauge theory in any dimension
with or without supersymmetry, whose gauge group
is $\prod_i U(N_i)$ with some bifundamental matter and consider the limit where $N_i>>1$.  We assume
the interactions involve a finite number of fields, which does not grow with $N_i$.  We consider a cousin theory
where we replace the gauge group with $\prod_i U(N_i+k_i|k_i)$ and the corresponding
bifundamental matter.  We assume the interactions involve traces of a fixed $N$-independent number
of fields in the theory.  
We consider the limit of this theory when $N_i>>1$.
We argue that as long as we consider gauge invariant fields made up of  less than $N_i$ fields:  {\it  there is a
1-1 correspondence between the correlation functions of these two theories to all orders in} $1/N$.  The proof
is simple\footnote{This idea was also used in the context of sigma models on supergroup manifolds in \cite{bvw}.}:
Consider the `t Hooft expansion of these two theories computing a correlation function of gauge invariant
operators involving a fixed number of fields which does not grow with $N_i$ (in particular we are not
considering Baryon type operators).  Consider correlation functions at a fixed order in $1/N$.  The two
theories will have exactly the same set of diagrams with the same Feynman amplitudes, except that
in one case the free loop index runs from $1,...,N_i$ and in the other one it runs over the ${\bf C}^{N_i+k_i|k_i}$
superspace index.  In the unitary case each such loop gives a factor of $N_i$ because $tr_i 1=N_i$.
In the supergroup case we have instead to take the supertrace.  This leads again to the factor
of $N_i$ for the i-th free index loop, because $Str_i 1=(N_i+k_i-k_i)=N_i$.    Thus to all orders in
the $1/N$ expansion the two theories yield identical correlation functions!\footnote{That there
is a natural correspondence between the operators in the $U(N+k|k)$ theory and $U(N+k-r|k-r)$ theory are known for WZW theories
on supergroups.  In such cases, there are fermionic nil-potent operators $Q_r$ in the $U(N+k|k)$ theory, which when we restrict
to their cohomology yield all the states in the $U(N+k-r|k-r)$ theory \cite{sch}.}

This raises the question of whether the two theories are identical.  From the viewpoint
of degrees of freedom, it is clear that $U(N_i+k_i|k_i)$ has more degrees of freedom
than its unitary cousin.  However one
may imagine that just as in the $1/N$ expansion there are cancellations between various
contributions and that there is no difference between the two.
There could potentially be two differences:  There could be differences between correlation
function of the two theories which are non-perturbative in $1/N$ expansion, say proportional to ${\rm exp}(-aN)$.
Furthermore there could be difference between correlation function of operators which involve $O(N)$ fields,
because our argument above was for operators with fixed number of fields.  In familiar large $N$ examples, these operators are expected  to have correlations which are non-perturbative in  $N$, i.e.,  $O({\rm exp}(-aN))$.  In particular one
can consider correlation functions of operators which vanish identically in the $U(N_i)$ theory (due to Casimir relations) but are
non-trivial in the $U(N_i+k_i|k_i)$ theory.
 We will give examples of this latter phenomenon in section 4.  
Thus the two theories are distinguishable as one would have expected.

What implications does this have for their gravity duals?
Let us consider two specific examples, though this idea applies to many other
known examples as well.  The considerations above apply in particular to the ${\cal N}=4$ supersymmetric theory
in 4 dimensions and the ABJM theory \cite{abjm}, where in each case we replace the group $U(N)$ by the supergroup $U(N+k|k)$.
As we have argued, to all orders in the $1/N$ expansion these theories are identical with their unitary cousins.
This in particular implies that the usual holographic duals for these theories, namely type IIB supergravity on $AdS^5\times S^5$
and M-theory on $AdS^4\times S^7$ work equally well for these theories.
 However as we have also argued,  there are differences between the unitary
and the non-unitary versions and so we expect different gravity duals which differ non-perturbatively by $O(exp(-a N))$
effects.  In the next section we will argue that this is indeed the case.

\section{Supermatrix models}

In this section we study supermatrix models.  Not only does this exemplify the general structure
we have discussed in section 3, but in fact more is true:  The computations
of supersymmetric partition functions often localize to integrals over eigenvalues of matrices.
In the context of partition functions of supersymmetric theories based on supergroups, the path-integral
localizes to integral over supermatrices.  So the lessons learned here will be directly
applicable to special amplitudes of the supersymmetric theories we have been discussing in section 3.
In fact, as we will discuss later in this section, the specific supermatrix amplitude we compute
can be interpreted as computing certain correlation functions in ${\cal N}=4$ SUSY theories in 4 dimensions
which distinguishes $U(N+k|k)$ theories from the $U(N)$ theory.

Hermitian matrix models at large $N$ have been thoroughly investigated over the years.  It is
natural to ask about their supergroup cousins.  In fact these were also studied early on \cite{yost,ag}.
It was found that to all orders in the $1/N$ expansion there is no difference between the two theories, consistent
with the general argument we outlined in the previous section.  An intuitive explanation of this was offered in
\cite{dv}, which we now review.

Consider matrix model associated to $U(N+k|k)$ gauge symmetry.  In the usual Hermitian case, when one
gauge fixes to a diagonal matrix, integrating the contribution of ghosts, coming
from the off diagonal gauge transformations leads to the Vandermonde of the diagonal eigenvalues.
These come from integrating out the ghost contributions to the action of the form
$$b_{ij}(x_i-x_j)c_{ij}\rightarrow \Delta (x_i)=\prod_{i<j} (x_i-x_j)^2$$
where $x_i$ denote the diagonal entries of the matrix.  The structure of the
Vandermonde is interpreted as eigenvalue repulsion.  In particular we can think of the Vandermonde as a 
$log(x_i-x_j)^2$ term in the action, which can be interpreted as a Coulomb repulsion potential
between same sign charges.
In the context of supergroups, the same manipulations work, except that we end up with superVandermonde.
What this means is that the off-diagonal elements which correspond to the fermionic directions
in the supergoup will now have bosonic ghosts associated with them, leading to determinant
factors in the denominator instead of the numerator.  If $(x_i,y_j)$ label the diagonal elements
of the $U(N+k|k)$ matrix (with $i=1,...,N+k$ and $j=1,..., k$, we will now have
$$\Delta(x_i,y_j)={\prod_{i<i', j<j'} (x_i-{x}_{i'})^2(y_j-y_{j'})^2\over \prod_{i,j} (x_i-y_j)^2} $$
This can also be interpreted in terms of associating `charges' to the eigenvalues, where
the first $N+k$ eigenvalues have $+$ charge and the last $k$ have a $-$ charge.  The numerator
can be interpreted as the repulsion of same sign charges and the denominator as the attractive
potential of opposite sign charges.  The statement that to all orders in $1/N$ the amplitudes of the $U(N)$
theory and the $U(N+k|k)$ theory are the same can be interpreted as the statement that $k$ of the negative charges will
screen $k$ of the positive charges leading to $N$ free positive charges thus leading to the same
answer as the $U(N)$ theory\footnote{This is similar to the intuitive explanation of
the relation between $U(N)$ Chern-Simons theory and $U(N+k|k)$ Chern-Simons theory \cite{va}.}. If we consider the matrix model with a polynomial
of degree $n$ potential $W(\Phi)$, then
the large $N$ dual involves a spectral curve given by \cite{dv2}
$$y^2=W'^2(x)+g(x)$$
where $g(x)$ is a polynomial of degree $n-2$ depending on how one distributes the eigenvalues among critical points.  The spectral curve is identical to the unitary case.  Here $ydx$ is interpreted as the eigenvalue density. For example
in the Gaussian case
$$W=Str (\Phi)^2$$
leads to spectral curve
$$y^2=\mu-x^2$$
where $\mu =N\lambda$.
  However one aspect of the supergroup case is nicer than the Hermitian case:
In the Hermitian case, the existence of the double cover, the fact that $y$ has an ambiguity and is
defined only up to a $+/-$ sign does not have a nice interpretation in the original theory and
is viewed as a large $N$ artefact \cite{maldaetal}.   However in the supergroup case, there are two sets
of eigenvalues and $ydx$ is interepreted as the `net' eigenvalue density.  Thus flipping
$$y\rightarrow -y$$ 
is exchanging the two sets.  In other words, now both $y$ and $-y$ are physical
and describe the geometry as seen from the vantage point of the upper or lower blocks
of the $U(N+k|k)$ matrix.

This picture explains why at large $N$ we expect the amplitudes
for matrix and supermatrix models to be the same.
 However it also suggests what we need to do to see
the difference between the two theories:  We need to `liberate' opposite pairs of charges from one another.  This
can be achieved if we have gauge invariant operators whose insertion in the path-integral can cancel the attractive force
of the opposite sign charges represented by canceling the denominator of the superVandermonde.
As already discussed in section 2, the denominator is a $U(N+k|k)$ invariant Casimir and so there is a `large'
gauge invariant operator ${\cal O}$
whose insertion will cancel the denominator and could potentially exhibit the difference between the unitary
and non-unitary theories.  This is analogous to considering analog of the `black hole background' in this theory where the pair can be created from vacuum,
leading to a non-unitary Schwinger type pair creation, which has been used as a model of Hawking's black
hole evaporation \cite{review}.  Insertion of this operator leads here to eigenvalue pair creation from vacuum
which is the source of the difference between non-unitary supermatrix models and unitary matrix models.
 We will now compute the effect of such an insertion.

For concreteness let us consider the case of $U(N+1|1)$ theory with a potential $W(\Phi)$.  In other words we consider the matrix integral
$$\int d\Phi\  {\rm exp}(-W(\Phi)/\lambda)\rightarrow \int dx_i dy {\Delta (x_i)\over \prod (x_i-y)^2}{\rm exp}(-\sum_i (W(x_i)-W(y))/\lambda)$$
Now consider the operator ${\cal O}$ which in the diagonal basis corresponds to $\prod (x_i-y)$.  As already discussed
in section 2 this corresponds to some combination of supertraces of degree $N+1$:
$${\cal O}=\sum_{\sum_i a_i=N+1} C_{\{a_i\}} \prod_i Str{\Phi}^{a_i}$$
This operator corresponds to a vanishing operator ${\cal O}=0$ in the $U(N)$ theory as already discussed in section 2, as 
is clear when we set $x_{N+1}=y=0$ and can be interpreted as a Casimir relation in that theory.  To exhibit
a difference between the $U(N)$ theory and $U(N+1|1)$ theory, it thus suffices to show that some correlations
involving ${\cal O}$ do not vanish in the $U(N+k|k)$ theory.

So now consider the correlation function
$$\langle {\cal O}{\cal O}\rangle={1\over Z}\int \prod_i (x_i-y)^2\cdot{
\Delta (x_i)\over \prod (x_i-y)^2}{\rm exp}(-\sum_i (W(x_i)-W(y))/\lambda)=$$
$$
={1\over Z}\int dx_i dy \Delta(x_i) {\rm exp}(-\sum_i (W(x_i)-W(y))/\lambda =$$
$$={1\over Z}\int dx_i \Delta(x_i) \ {\rm exp}(-\sum_i W(x_i)/\lambda)\  .\  \int dy \ {\rm exp}(W(y))/\lambda )$$
We see that the two sets of eigenvalues have been decoupled.  We now estimate
this amplitude at large $N$.  Let $Z_N$ denote the partition function of the $U(N)$ theory.
We know that this scales at large $N$ as $exp(-a N^2+O(1))$.  The partition function of the $U(N+k|k)$ theory
$Z$ as already discussed to leading order is also give by that of the $U(N)$ theory and so it also scales
the same way at large $N$.  The numerator is a product of two factors:  The $x$-integral is the
same as the partition function of a $U(N+1)$ Hermitian matrix model and so its amplitude will
scale as $exp(-a(N+1)^2+O(1))$ and the integral over the $y$ of order $O(1)$.
Thus we find that the correlation scales as
$$\langle {\cal O}{\cal O}\rangle \sim A \ {exp(-a(N+1)^2)\over exp(-a N^2)}\sim A'\  exp(-2a N)$$
(where A is a non-zero constant).  We have thus shown that the correlation functions of some operators in the
$U(N+1|1)$ theory and $U(N)$ theory differ by non-perturbative terms in the $1/N$ expansion.\footnote{The striking analogy between the matrix model and black holes may not be accidental.  In fact the
Gaussian matrix model in the $\beta$-ensemble corresponds to non-critical $c=1$ matrix model
on a circle of radius $1/\beta$ \cite{dv3}.  So the NS limit \cite{ns} of this theory where $\beta \rightarrow 0$ should
correspond to non-critical non-compact c=1 theory which has been argued to be related to questions
involving scatterings in the gravitational 2d theory \cite{pol,newhat}.}

\subsection{Amplitudes on $S^4$ for ${\cal N}=4$ SYM}
In this section we show how the computations done in the context of supermatrix models can be used
to show that $U(N+k|k)$ is distinguishable from the $U(N)$ theory non-perturbatively.  For simplicity
we focus on the $k=1$ case, i.e. $U(N+1|1)$ gauge theory.  Consider putting both theories on $S^4$
as was done in \cite{pestun1,pestun2}.  As was shown there the computation for supersymmetric Wilson loops reduces to an integrals
over the eigenvalues of the $U(N)$ matrix.  The same reasoning shows that also for chiral operators 
$tr \phi^k$ at the north
pole, the computation of the amplitude reduces to
$$\langle tr \phi^k \rangle ={1\over Z}\int da_i \prod_{i<j} (a_i-a_j)^2\ \big(\sum_i a_i^k\big) \ {\rm exp}\big[{-1\over \lambda}\sum_i a_i^2\big]$$
where
$$Z=\int da_i \prod_{i<j} (a_i-a_j)^2\  {\rm exp}\big[{-1\over \lambda}\sum_i a_i^2\big]$$
$${1\over \lambda }={8\pi^2 r^2\over g_{YM}^2}$$
The ideas of localization work equally well for the supergroups and in that case
the only difference is that instead of traces we get supertraces.
 Consider the operator ${\cal O}$ of the $U(N+1|1)$ theory which is an order $N+1$ Casimir
 of the chiral field $\phi$, which is the invariant Casimir associated with
 $${\cal O}_{N+1}\leftrightarrow \prod_{i=1}^{N+1} (x_i-y) =\sum a Str (\phi^{N+1})+...$$
which was already discussed.  Note that in the $U(N)$ theory it is identically zero and we thus have
$$\langle {\cal O}_{N+1}^2\rangle_{U(N)}=0$$
However in the $U(N+1|1)$ theory we obtain, as already discussed in the context of supermatrix models:
$$\langle {\cal O}_{N+1}^2\rangle_{U(N+1|1)}\sim A \ {\rm exp}(-aN)\not= 0$$
This computation shows that non-perturbatively the ${\cal N}=4$ SYM with $U(N+1|1)$ and $U(N)$ are distinguished
by non-perturbative effects in $1/N$.

\section{Concluding remarks}
In this paper we have argued for the extension of holography to the non-unitary setup.
We have assumed that non-unitary conformal theories of the type we have discussed exist.
Note that there already is strong evidence that non-unitary conformal theories
do exist in dimensions $2\leq d \leq 6$.  In particular the Yang-Lee edge singularity
(the fixed point for real scalar field with $ig\phi^3$ interaction) is believed to exist
in $2\leq d \leq 6$.  Moreover the dimension of the basic field in the theory is non-trivial
and computed by a number of methods, including the recent bootstrap method, leading
to consistent results (see \cite{boot} and references therein).  In this paper we have been assuming in particular that the $U(N+k|k)$ theory
with ${\cal N}=4$ supersymmetry exists in 4 dimensions.
Even though this is reasonable to believe, given the high degree of supersymmetry, the main point
of this paper, which is the existence of non-unitary holography can probably be established using
the known 2d non-unitary models which should lead to non-unitary $AdS^3$ holographic duals
(see \cite{log} for concrete proposals along this line).

We have provided evidence that familiar backgrounds such as $AdS^5\times S^5$
admit non-unitary non-perturbative completions.  Nevertheless, it still seems true that there is a unique
unitary completion of it.  However, even though we may `prefer' to live in such a unitary universe,
it seems  plausible that logical consistency is not what picks that out.  In other words, {\it unitarity 
is not a necessary consequence of holography}.

Turning this around, and assuming we indeed live in a non-unitary universe, we may have found a way
to reconcile Hawking's original perspective of non-unitarity of black hole evaporation \cite{hawking} with holography.
Of course there is no guarantee that the non-unitarity of the theories we have been discussing
are of the same type needed in the context of Hawking's original proposal for non-unitarity of black hole evaporation. 
It would be interesting to try to flesh out how this works in the context of non-unitary holography and
a good toy model may be the non-critical superstring studied in \cite{newhat}.  This
may provide a way of reconciling the firewall paradox \cite{amps} with holography.   It could also
provide a way to resolve the AdS/CFT wormhole paradox raised in \cite{jn}.

One can also ask whether this leads to a non-unitary non-perturbative completion of string theory.
For example the $N\rightarrow \infty$ limit will give a theory essentially in flat space but with
non-unitary non-perturbative completion.
Since holography can be taken as a non-perturbative definition of string theory in certain backgrounds,
this would then imply that string theory does admit such non-unitary non-perturbative completions.
This would suggest that for example D3 branes with $U(1|1)$ gauge symmetry exist in this
version of string theory, and our ${\cal N}=4$ supersymmetric $U(N+k|k)$ theory can be viewed as the theory living on
$N$ ordinary D3 branes and $k$ D3 branes of type $(1|1)$ brought together.
It would be interesting to uncover the structure of such non-unitary completions of string theory.
A particularly interesting limit to consider is the limit $k\rightarrow \infty$.  This would
correspond to the holographic dual of $U(N+\infty|\infty)$, which is a maximally non-unitary
version of the theory in the opposite extreme from the unitary $U(N)$ theory.

\section*{Note added}
After this paper was submitted, it was pointed out to me\footnote{I thank B. Heidenreich and M. Reece for pointing
this out.}\ that branes of the type needed for this paper have already been introduced in the context of
superstrings in a beautiful paper \cite{taka}.  This paper uses the same strategy for obtaining supergroups
as was done in \cite{va} to obtain supergroups in topological strings, namely simply adding a minus sign
to the $(0|1) $ type brane boundary states.  They have also studied the effect of introducing
these branes on the duality web and have argued that the string dualities seem to work also incorporating these new branes.
Their conclusion that the $U(N+k|k)$ theory and $U(N)$ theory are identical perturbatively is in
agreement with what we have argued here.  However, as we have also argued in this paper, unlike their claim, the two theories are not
 identical due to distinct Casimir relations, leading to different operator algebras, with computable $O(exp(-aN))$ effects.

\section*{Acknowledgements}

 I would like to thank H. Arfaei, N. Berkovits, R. Dijkgraaf, B. Haghighat, D. Jafferis, J. Maldacena, S. Mathur, N. Nekrasov, H. Ooguri, A. Polyakov,
 S. Rey, M. Rocek, S. Sachdev, V. Schomerus, M. Soroush, A. Strominger, E. Witten and X. Yin for valuable discussions.  In addition I would like
to thank SCGP for hospitality during the 12th Simons workshop in Mathematics and Physics, where
this project was initiated.

This research is supported in part by NSF grant PHY-1067976.

\bibliography{references}

\end{document}